\documentclass[%
reprint,
amsmath,amssymb,
aps,
]{revtex4-2}

\usepackage{graphicx}
\usepackage{dcolumn}
\usepackage{bm}

\usepackage{makecell}
\usepackage{caption}
\usepackage{subcaption}
\usepackage[colorlinks,linkcolor=blue,citecolor=blue,urlcolor=blue,bookmarks=false,hypertexnames=true,linktocpage=true]{hyperref}
\usepackage{xcolor}

\begin{document}
    \preprint{APS/123-QED}
	
 \title{Quadrature-witness readout for backscatter mitigation in gravitational-wave detectors limited by back-action}
    \author{Niels B\"ottner}
    \author{Roman Schnabel}
    \author{Mikhail Korobko}
    \email{mikhail.korobko@uni-hamburg.de}
    \affiliation{Institut f\"ur Quantenphysik und Zentrum für Optische Quantentechnologien , Universit\"at Hamburg, Luruper Chaussee 149, 22761 Hamburg, Germany}
	
    \date{\today}
	
    \begin{abstract}
        Disturbances in gravitational wave (GW) observational data are often caused by non-stationary noise in the detector itself, such as back-scattering of laser stray light into the signal field. Unlike GW signals, non-stationary noise can appear in both the GW-signal quadrature and the orthogonal quadrature, which is usually not measured. Simultaneous sensing of this orthogonal quadrature provides a witness channel that can be used to reconstruct the disturbance in the signal quadrature enabling a subtraction of non-stationary noise. Here, we present the concept of quadrature witness that is compatible with frequency-dependent squeezing, which is already used to simultaneously reduce photon shot noise and photon radiation pressure noise. We demonstrate that implementing this approach in a GW detector could reduce noise caused by loud back-scatter events, thereby improving the overall sensitivity and robustness of GW observatories.
    \end{abstract}
    
    \maketitle

\section{\label{sec:Introduction}Introduction}
Since the first observation of a gravitational wave (GW) in 2015 by the two LIGO detectors~\cite{Abbott2016}, over 200 GW events from binary mergers have been observed. While the current generation of GW detectors observes new events every few days~\cite{TheLIGOScientificCollaboration2017a,TheLIGOScientificCollaboration2019,LIGO2022O31st,LIGO2023O32nd,abac2025gwtc}, future detectors, such as the Einstein Telescope~\cite{Branchesi2023} and the Cosmic Explorer~\cite{Evans2021} will bring this rate to several events per minute.
Such an improvement in the detection capabilities will bring unprecedented progress in scientific understanding of the Universe~\cite{abac2025science}, but it will require significant technological advancement.
The sensitivity of the current observatories, Advanced LIGO~\cite{AdvancedLIGO15}, Advanced Virgo~\cite{Acernese2020}, and KAGRA~\cite{Akutsu2019KAGRA}, 
is limited by various sources of noise. One of the most fundamental noises is quantum: 
it originates from the ground state fluctuations of the quantum vacuum field entering the output port of the interferometer. 
These fluctuations couple to the amplitude and phase of the light field inside the detector, leading to quantum radiation-pressure noise (QRPN) on the mirrors and photon shot noise (SN) upon detection. 
SN is the main limitation of detector sensitivity at mid frequencies and above ($\gtrsim100\,\text{Hz}$), while QRPN is one of the limiting noises at lower frequencies ($\lesssim30\,\text{Hz}$). 
Modern detectors employ quantum squeezed light to suppress quantum noise\,\cite{Caves1981, Schnabel2010, Schnabel2017}.
Since its first implementation in the GEO\,600 GW detector in 2011\,\cite{LSC2011,Grote2013}, squeezed light has been brought to continuous operation in both Advanced LIGO and Advanced Virgo~\cite{Tse2019,AcerneseaVIRGOsqueezing2019}, culminating in the simultaneous reduction of both SN and QRPN in Advanced LIGO~\cite{Jia2024} using a frequency-dependent squeezing approach~\cite{Unruh1983, JaekelQuantumLimits1990, Kimble2001, ChelkowskiFDSexp2005}. 
Future detectors will reach even higher quantum enhancement employing advanced quantum technology~\,\cite{danilishin2012quantum, Danilishin2019, korobko2025quantum}.

Quantum noise is not the only factor limiting the detector sensitivity. Back-scattered laser light is a dominant, non-stationary noise source: tiny fractions of laser light from the arm cavities of the detectors scatter out of the main beam, reflect of moving surfaces and then re-enter the main beam, imprinting surface motion as a parasitic phase/amplitude modulation that can mask the genuine GW signal~\cite{Vinetscatteredlight1996,Vinetscatteredlight1997,Vahlbruch2007,Ottawayscatteredlight2012}.  
Back-scatter noise is commonly mitigated by using higher-quality optical surfaces to reduce scattering, blackening potential back-scatter surfaces, reducing the motion of the back-scatter surfaces relative to the mirrors of the arm resonators~\cite{Vahlbruch2007}, precise modeling of the effects~\cite{soni2024modeling} and baffles to intercept and absorb scattered light~\cite{CarcasonaScatteredLight2023}.
These measures remain imperfect and further suppression is required.

Quantum technology provides a complimentary path by enabling direct readout of these disturbances and their coupling to GW signal.
Quantum dense metrology (QDM) uses two-mode squeezed light to simultaneously read out two non-commuting field quadratures with a precision better than photon SN~\cite{SteinlechnerS2013}.
Because the back-scattered disturbance couples both into the signal quadrature and in the orthogonal quadrature (which contains no GW signal), simultaneously detecting both allows the orthogonal quadrature to serve as a witness for subtraction of the disturbance from the signal. 
QDM enables this approach enhanced with squeezed-light injection\,\cite{AstM2016}.
However, neither this quantum approach nor similar non-squeezed techniques~\cite{MeindersScattering2015,voigt2023simulating,voigt2025tunable,Lohde2025}, have been extended to interferometers operating in the back-action-limited regime at low frequencies -- precisely where both back-scattered light and QRPN are most significant.

Here we provide that extension.
We present a new quantum-enhanced readout that preserves frequency-dependent squeezing of the signal quadrature while extracting the back-scatter witness from the orthogonal quadrature.
This scheme employs two squeeze lasers to enable the witness readout without sacrificing the high level of frequency-dependent quantum enhancement for the signal quadrature. We perform a statistical analysis of random scattering events and show that already in the near-term upgrades of existing detectors, a non-invasive implementation can remove the loudest scatter events with only minimal loss of GW sensitivity -- making it a promising approach for detectors affected by the back-scatter noise.\\

\section{Quadrature-witness readout}
The main idea of the dual readout is to split the signal that comes out of the detector onto two independent balanced homodyne detectors (BHDs), one of which measures the GW signal (together with back-scatter disturbance) and the other one -- only the back-scatter contribution. However, this split creates a loss channel for the squeezed light, where a quantum vacuum state couples into the squeezed state and leads to severe decoherence\,\cite{Schnabel2017}.
Quantum-dense metrology suggests to replace this quantum vacuum state by an EPR-entangled state, negating the detrimental effects of loss. However, combining it with frequency-dependent squeezing requires employing sophisticated quantum system, as we show in the companion paper [REF].
Instead, in this work, we propose replacing this vacuum field with another squeezed field, which we will call readout squeezing. This also negates the detrimental effects of splitting the signal, but prevents achieving quantum enhancement on witness quadrature, as done in QDM.

In this section, we start by presenting the general concept of our dual readout approach and its theoretical foundation. We demonstrate its compatibility with frequency-dependent squeezing and provide an intuitive theoretical description that exemplifies the resulting quadrature sensitivities, and outline the optimization procedure that allows to achieve high sensitivity in a realistic setting.
    
    \subsection{Setup and sensitivities\label{sec:setup:sens}}

     \begin{figure*}
		  \includegraphics[width=6.75in]{"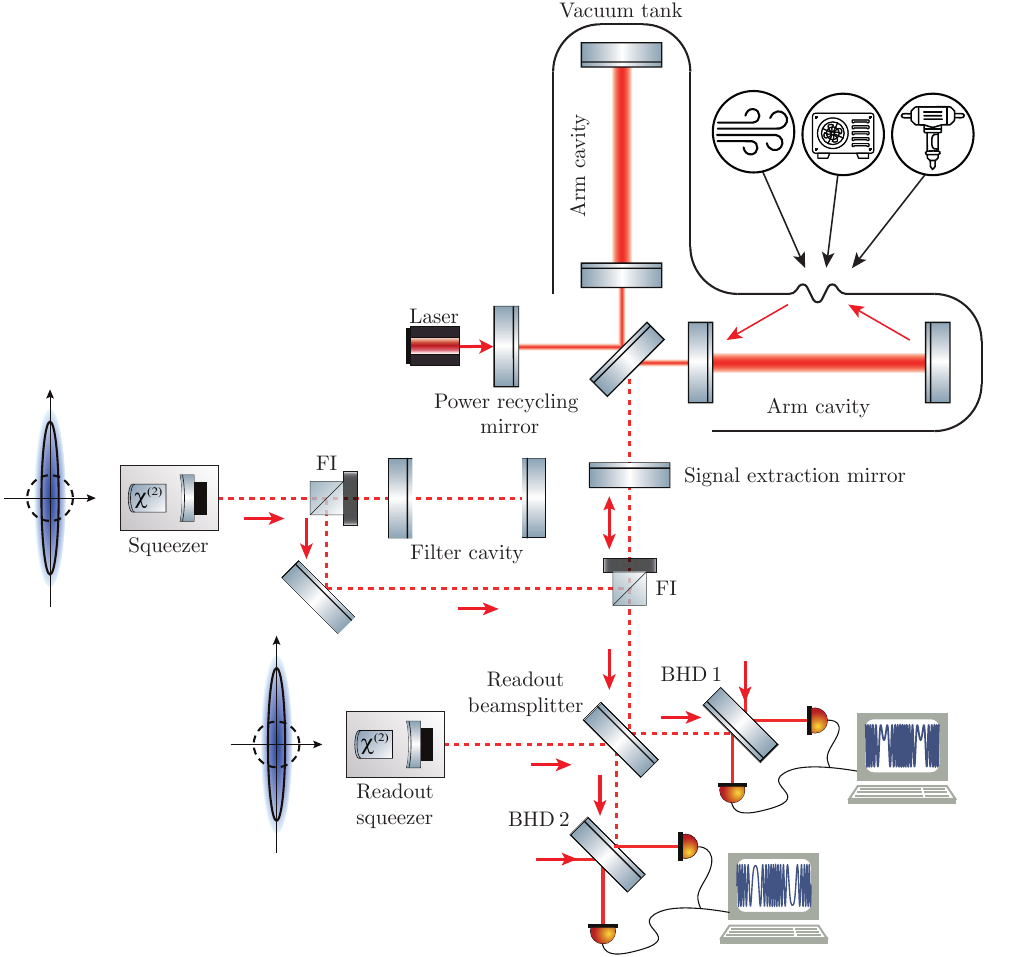"}
		  \caption{\label{fig:setup}Conceptual setup of our approach in a GW detector. The walls of the vacuum tanks are excited by  microseims and anthropogenic sources. Dual readout is implemented by the readout beamsplitter and two balanced homodyne detectors (BHD) in the output of the interferometer. One BHD measures the signal quadrature while the other one measures the witness quadrature. A second squeezed state in injected through the open port of the readout beamsplitter. Frequency-dependent squeezing is realized by a filter cavity and injected into the detector through a Faraday isolator (FI).}
	\end{figure*}
    \noindent

    The detector setup, shown in Fig.~\ref{fig:setup}, consists of three core components: a Michelson interferometer with arm, power recycling and signal extraction cavities; the injection of frequency-dependent squeezed light; and the dual readout. Frequency-dependent squeezing is achieved by reflecting squeezed light off a detuned filter cavity. This creates a frequency-dependent rotation of the squeezed state of light\,\cite{Kimble2001}. The reflected squeezed light is then coupled into the detector through a Faraday isolator. Dual readout is realized by a readout beamsplitter, which sends the signal onto two balanced homodyne detectors. One of them measures the signal quadrature, while the other one -- orthogonal witness quadrature. As a novelty of this work, we inject squeezed light into the detector through the open port of the readout beamsplitter. The purpose of doing this is to avoid vacuum entering the detector through this port which would reduce the amount of detected squeezing. In this subsection, we consider the simple case of a balanced readout beamsplitter. 
    Equal splitting leads to the loss of $50\%$ of the signal in both quadratures, resulting in $3\,\text{dB}$ penalty to the sensitivity at high frequencies. In the next sections, we relax this condition to further optimize the sensitivity.
    
\begin{figure}
 		\begin{subfigure}{1\linewidth}
			\includegraphics[width=\textwidth]{"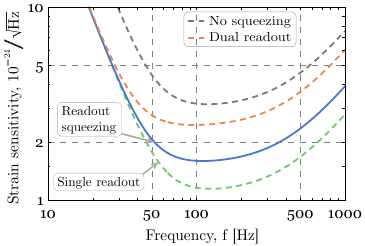"}
            \phantomsubcaption
                \vspace{-0.35cm}
		\end{subfigure}
		\begin{subfigure}{1\linewidth}
			\includegraphics[width=\textwidth]{"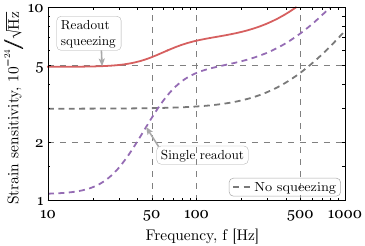"}
            \phantomsubcaption
		\end{subfigure}
		\caption{\label{fig:phaseampsym}Strain sensitivity of the signal quadrature (top) demonstrates that frequency-dependent squeezing is compatible with dual readout. 
        Quadrature-witness with readout squeezing (solid blue) maintains quantum enhancement at all frequencies compared to the case without squeezing (dashed grey), only suffering a 3\,dB penalty for dual readout at high frequency. Readout squeezing allows to restore a significant portion of sensitivity compared to the dual readout without readout squeezing (orange dashed).
        The witness quadrature sensitivity (bottom) is sufficient to allow effective back-scatter subtraction. The detector parameters are chosen such that we achieve optimal frequency-dependent squeezing and reach approximately the same signal quadrature sensitivity as A+ for single readout (see Table \ref{tab:parameters:Aplus:sym}).}
	\end{figure}

    Fig.~\ref{fig:phaseampsym} shows the signal and witness quadrature sensitivities of our approach. The parameters of the detector and the filter cavity were chosen such that we achieve optimal frequency-dependent squeezing and get approximately the same signal quadrature sensitivity as A+~\cite{Barsotti2018} for the single readout (see Table~\ref{tab:parameters:Aplus:sym}). At low frequencies, QPRN dominates the sensitivity of the signal quadrature, which is almost identical for single readout and for quadrature-witness with readout squeezing. In the latter case, both QRPN and signal are reduced by the readout beamsplitter, and their noise-to-signal ratio remains almost unchanged. At high frequencies, the sensitivity is limited by SN, which is not affected by the readout beam-splitter (without squeezing), since the noise lost is compensated by the added vacuum field. Half of the signal is still lost, and thus noise-to-signal ratio is reduced by 50\%. In this case, replacing the vacuum field entering the readout beamsplitter with squeezed vacuum significantly increases the signal quadrature sensitivity compared to dual readout.
    
    The witness quadrature sensitivity has more complex frequency-dependent behavior. In a single readout, the witness quadrature is only accessible if the readout quadrature is switched away from the signal, but it is useful to compute its sensitivity as a reference.
    In this case, the quadrature of the squeezed state is optimized in a frequency-dependent way to suppress QRPN at low frequencies and SN at high frequencies. Therefore, the witness quadrature is highly squeezed at low frequencies and anti-squeezed at high frequencies. For the quadrature-witness readout, the phase of the readout squeezing is aligned with squeezing in the signal quadrature. Thus, the witness quadrature has two contributions: squeezed from the main squeezing, and anti-squeezed from the readout squeezing. At high frequency, main squeezing turns to anti-squeezed quadrature due to frequency-dependent squeezing, which further increases noise in the witness quadrature.
    Similarly to the case above, the loss of 50\% of the signal leads to the $3\,\text{dB}$ reduction in the sensitivity at high frequency.
    Despite increased noise for the witness quadrature, its measurement still allows to subtract a significant fraction of back-scatter light, which appears predominantly at low frequencies.

    \begin{table}[h]
		\caption{\label{tab:parameters:Aplus:sym} Parameters used to get the sensitivity close to the one of A+~\cite{Barsotti2018}.}
		\begin{ruledtabular}
			\begin{tabular}{lcccc}
				Parameter & Value 
				\\
				\hline
				Laser wavelength 											            & $1064\,\text{nm}$\\
				Arm circulating power 										            & $750\,\text{kW}$\\
				Test mass weight 											            & $40\,\text{kg}$\\
				Input test mass power transmission 							            & $1.4\%$\\
				End test mass power transmission 							            & $5\,\text{ppm}$\\
				Signal extraction mirror power transmission 				            & $32.5\%$\\
				Arm cavity length 											            & $3995\,\text{m}$\\
				Signal extraction cavity length 							            & $55\,\text{m}$\\
                \makecell[l]{Single pass loss inside the\\ signal extraction cavity}    & $1000\,\text{ppm}$\\
				Total detection efficiency 							                    & $90\%$\\
				Squeezer: injected external squeezing 						            & $7.5\,\text{dB}$\\
                Readout squeezer: injected external squeezing 						    & $7.5\,\text{dB}$\\
                Readout beamsplitter power reflectivity 						        & $50\%$\\
			\end{tabular}
		\end{ruledtabular}
	\end{table}
    
    \subsection{Theory}

    Here, we follow the formalism of Ref.~\cite{danilishin2012quantum} to derive simplified expressions for the signal and witness quadrature sensitivities. We start with a Michelson interferometer without a signal recycling cavity. The output is described by:
    \begin{equation}\label{eq:basicIFO}
            \begin{pmatrix}
                b^{w}\\
                b^{s} 
            \end{pmatrix}= 
        e^{2i\phi} \mathbb{M} a_{\text{in}}+e^{i\phi}
            \begin{pmatrix}
                G^{w}x^{w}\\
                G^{s}x^{s} 
            \end{pmatrix}
    \end{equation}
    \noindent
    with
    \begin{equation}
        \mathbb{M}=
            \begin{pmatrix}
                1 & 0\\
                -\mathcal{K}(\Omega) & 1
            \end{pmatrix},
    \end{equation}
    \noindent
    and input
    \begin{equation}
        a_{\text{in}}=
             \begin{pmatrix}
                 a_{\text{in}}^{w}\\
                 a_{\text{in}}^{s}
            \end{pmatrix}.
    \end{equation}
    Here, $b^{w}$ is the output in the witness quadrature; $b^{s}$ is the output in the signal quadrature; $\mathcal{K}(\Omega)$ is the Kimble factor\,\cite{Kimble2001}; $\phi=\phi(\Omega)$ is the single trip phase of the interferometer; $G^{w}$ and $G^{s}$ are the witness and signal quadrature transfer functions of the normalized signals $x^{w}$ and $x^{s}$.
    Frequency-dependent squeezing is achieved by reflecting the input off a filter cavity before it is injected into the interferometer. In the lossless case, the reflection off a filter cavity is described by:
    \begin{equation}
        \mathbb{F}=\mathbb{P}[\theta(\Omega)] e^{i \beta},
    \end{equation}
    \noindent
    with the rotation matrix
    \begin{equation}
        \mathbb{P}[\theta(\Omega)]=
            \begin{pmatrix}
                \cos \theta(\Omega) & -\sin \theta(\Omega)\\
                \sin \theta(\Omega) & \cos \theta(\Omega)
            \end{pmatrix}.
    \end{equation}
    The filter cavity rotates the input state by the frequency-dependent rotation angle $\theta(\Omega)$ and adds a phase $\beta=\beta(\Omega)$. After adding frequency-dependent squeezing, we obtain:
    \begin{equation}
            \begin{pmatrix}
                b^{w}\\
                b^{s} 
            \end{pmatrix}= 
        e^{i(2\phi+\beta)} \mathbb{M} \mathbb{P}[\theta(\Omega)] a_{\text{in}}+e^{i\phi}
            \begin{pmatrix}
                G^{w}x^{w}\\
                G^{s}x^{s} 
            \end{pmatrix}.
    \end{equation}
    From this equation, we can compute the phase quadrature output for a standard detector with single readout:
    \begin{equation}\label{eq:singleOutput}
        b^{s} = e^{i(2\phi+\beta)}\sqrt{1+\mathcal{K}^{2}(\Omega)}a_{\text{in}}^{s}+e^{i\phi}G^{s}x^{s}
    \end{equation}
    \noindent
    where we set the rotation angle to $\theta(\Omega)=\arctan{\mathcal{K}(\Omega)}$. In this case, the noise-to-signal ratio is given by:
    \begin{equation}\label{eq:SNRsinglereadout}
        S_{h}^{s} = \dfrac{1+\mathcal{K}^{2}(\Omega)}{\lvert G^{s}\rvert^{2}}S_{\text{in}}^{s}
    \end{equation}
    \noindent
    with the signal quadrature spectral density of the input state $S_{\text{in}}^{s}$.\\
    To readout both quadratures simultaneously, we need to place a beamsplitter in the output port of the interferometer. Here, we consider the simple case of a balanced beamsplitter. In the case of dual readout, Eq.\,\ref{eq:singleOutput} changes to:
    \begin{equation}
            \begin{pmatrix}
                b^{w}\\
                b^{s} 
            \end{pmatrix}= 
        \dfrac{1}{\sqrt{2}}\left[e^{i(2\phi+\beta)} \mathbb{M} \mathbb{P}[\theta(\Omega)] a_{\text{in}}+e^{i\phi}
            \begin{pmatrix}
                G^{w}x^{w}\\
                G^{s}x^{s} 
            \end{pmatrix}\right]
            +\dfrac{1}{\sqrt{2}} v_{\text{BS}}
    \end{equation}
    \noindent
    with vacuum
    \begin{equation}
        v_{\text{BS}} =
            \begin{pmatrix}
                v^{w}\\
                v^{s}
            \end{pmatrix}
    \end{equation}
    \noindent
    that couples into the detector through the open port of the beamsplitter. This vacuum limits the sensitivity of the detector. Therefore, we replaced it with a second input
    \begin{equation}
        a_{\text{BS}}=
             \begin{pmatrix}
                 a_{\text{BS}}^{w}\\
                 a_{\text{BS}}^{s}
            \end{pmatrix},
    \end{equation}
    \noindent
    which could have a squeezed signal or witness quadrature:
    \begin{equation}
            \begin{pmatrix}
                b^{w}\\
                b^{s} 
            \end{pmatrix}= 
        \dfrac{1}{\sqrt{2}}\left[e^{i(2\phi+\beta)} \mathbb{M} \mathbb{P}[\theta(\Omega)] a_{\text{in}}+e^{i\phi}
            \begin{pmatrix}
                G^{w}x^{w}\\
                G^{s}x^{s} 
            \end{pmatrix}\right]
            +\dfrac{1}{\sqrt{2}} a_{\text{BS}}.
    \end{equation}
    \noindent
    As before, we compute signal quadrature output:
    \begin{equation}\label{eq:dualOutput}
        b^{s} = \dfrac{1}{\sqrt{2}}\left[e^{i(2\phi+\beta)}\sqrt{1+\mathcal{K}^{2}(\Omega)}a_{\text{in}}^{s}+e^{i\phi}G^{s}x^{s}+a_{\text{BS}}^{s}\right]
    \end{equation}
    \noindent
    and the spectral density of the output:
    \begin{equation}\label{eq:dualOutputPSD}
        S^{s}_{\text{out}}=\dfrac{1}{2}\left[\left(1+\mathcal{K}^{2}(\Omega)\right)S_{\text{in}}^{s}+\lvert G^{s}\rvert^{2}+S^{s}_{\text{BS}}\right]
    \end{equation}
    \noindent where $S^{s}_{\text{BS}}$ is the signal quadrature spectral density of the second input state $a_{\text{BS}}$.\\
    If we consider the case where $a_{\text{in}}$ and $a_{\text{BS}}$ both have squeezed signal quadratures with identical spectral densities $S_{\text{SQZ}}$, we obtain the following noise-to-signal ratio:
    \begin{equation}
         S_{h}^{s} = \dfrac{2+\mathcal{K}^{2}(\Omega)}{\lvert G^{s}\rvert^{2}}S_{\text{SQZ}}^{s}.
    \end{equation}
    At low frequencies, the $\mathcal{K}^{2}(\Omega)$ term dominates and we obtain the same sensitivity as for single readout (see  Eq.\,\ref{eq:SNRsinglereadout}). At higher frequencies, the $\mathcal{K}^{2}(\Omega)$ term becomes negligible and the sensitivity is reduced by a factor of two compared to single readout.\\
    In the same way, we can compute the witness quadrature output:
    \begin{equation}\label{eq:dualOutputAmp}
        b^{w} = \dfrac{1}{\sqrt{2}}\left[\dfrac{e^{i(2\phi+\beta)}}{\sqrt{1+\mathcal{K}^{2}
        (\Omega)}}\left(a_{\text{in}}^{w}-\mathcal{K}(\Omega)a_{\text{in}}^{s}\right)
        +e^{i\phi}G^{w}x^{w}
        -a_{\text{BS}}^{w}\right]
    \end{equation}
    \noindent
    and the corresponding spectral density:
   \begin{equation}\label{eq:dualOutputPSDAmp}
        S^{w}_{\text{out}}=\dfrac{1}{2}\left[\dfrac{S_{\text{in}}^{w}}{1+\mathcal{K}^{2}
        (\Omega)}+\dfrac{\mathcal{K}^{2}(\Omega)S_{\text{in}}^{s}}{1+\mathcal{K}^{2}
        (\Omega)}+\lvert G^{w}\rvert^{2}+S^{w}_{\text{BS}}\right]
    \end{equation}
    \noindent
    where $S^{w}_{\text{in}}$ and $S^{w}_{\text{BS}}$ are the witness quadrature spectral densities of the input states $a_{\text{in}}$ and $a_{\text{BS}}$, respectively.\\
    As before, we assume that $a_{\text{in}}$ and $a_{\text{BS}}$ both have squeezed signal quadratures with identical spectral densities $S_{\text{SQZ}}$. The noise-to-signal ratio is as follows:
    \begin{equation}\label{eq:dualOutputSNRAmp}
         S_{h}^{w} = \dfrac{2+\mathcal{K}^{2}(\Omega)}{\left(1+\mathcal{K}^{2}
        (\Omega)\right)\lvert G^{w}\rvert^{2}}S_{\text{SQZ}}^{w}+\dfrac{\mathcal{K}^{2}(\Omega)}{\left(1+\mathcal{K}^{2}
        (\Omega)\right)\lvert G^{w}\rvert^{2}}S_{\text{SQZ}}^{s}.
    \end{equation}
    \noindent
    At low frequencies, $\mathcal{K}(\Omega)$ is very large and Eq.\,\ref{eq:dualOutputSNRAmp} simplifies to:
    \begin{equation}
         S_{h}^{w} \approx \dfrac{S_{\text{SQZ}}^{w}+S_{\text{SQZ}}^{s}}{\lvert G^{w}\rvert^{2}}\approx\dfrac{S_{\text{SQZ}}^{w}}{\lvert G^{w}\rvert^{2}}.
    \end{equation}
    \noindent
    Since we consider squeezed states in signal quadrature, the witness quadrature component of the input states is anti-squeezed and so the sensitivity.\\
    At high frequencies, $\mathcal{K}(\Omega)$ becomes very small and the sensitivity is given by;
    \begin{equation}
         S_{h}^{w} \approx \dfrac{2S_{\text{SQZ}}^{w}}{\lvert G^{w}\rvert^{2}}.
    \end{equation}
    \noindent
    Again we can see the reduction in sensitivity by a factor of two due to dual readout.

    \subsection{Optimization}
    The two main parameters introduced by the quadrature-witness readout are the level of readout squeezing and the transmission of the readout beam-splitter. In the previous section, we assumed both to be fixed: beam splitter to be 50/50, and the squeeze value to match that of the main squeezing. They can be optimized for better performance. In particular, the main squeeze value is defined by the total loss in the GW detector and is limited to a moderate value to reduce the impact on the QRPN\,\cite{korobko2025quantum}. Readout squeezing only suffers the readout loss (typically, the inefficiency of the photodiodes and the imperfect mode overlap at the homodyne detector), which are comparatively small. Therefore, a much higher readout squeeze value can be used. Balancing between the signal lost at the readout beam-splitter, anti-squeezing introduced in the witness quadrature and squeezing restored to the signal quadrature, we find an optimal configuration that preserves high signal quadrature and increases the sensitivity in the witness quadrature.

\begin{figure}
		\begin{subfigure}{1\linewidth}
			\includegraphics[width=\textwidth]{"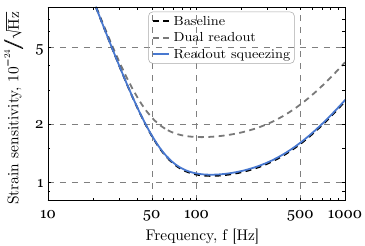"}
            \phantomsubcaption
                \vspace{-0.35cm}
		\end{subfigure}
		\begin{subfigure}{1\linewidth}
			\includegraphics[width=\textwidth]{"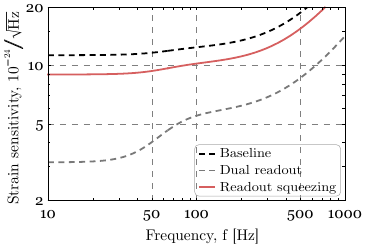"}
            \phantomsubcaption
		\end{subfigure}
		\caption{\label{fig:opt}Optimized signal (top) and witness (bottom) quadrature sensitivities. The baseline curves show the sensitivities for a detector with readout losses of $2\%$ after the readout beam-splitter and $15\,\text{dB}$ of readout squeezing. The solid curves show the sensitivities for the optimized readout beamsplitter power reflectivity of $69\%$ and amount of readout squeezing of $9.5\,\text{dB}$. For comparison, we plotted the sensitivity of the dual readout without readout squeezing for the same reflectivity.}
	\end{figure}

    In Fig.~\ref{fig:opt} we show an example of such optimization, considering 2\% of the readout loss after the readout beam splitter. The signal quadrature remains unaffected by the changes to the beam-splitter ratio from 50/50 (baseline) to 69/31 (detecting more signal quadrature), and the reduction of the readout squeeze value from 15\,dB to 9.5\,dB. Further optimization can be performed for more detailed loss budget for each specific detector design.

    \section{Back-scatter light signal and statistic}

    In this section, we present the details of the model that we use to generate scattering events. We also describe how we use this model to simulate a set of scattering events with a statistic that corresponds to those observed in GW detectors. We consider the sensitivity of the signal and witness quadratures as presented in Section~\ref{sec:setup:sens} without further optimization.

    \subsection{Model}

\begin{figure}
 		\begin{subfigure}{1\linewidth}
			\includegraphics[width=\textwidth]{"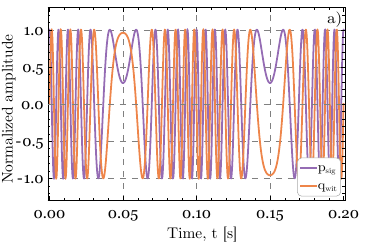"}
            \phantomsubcaption
            \label{fig:sc:time}
            \vspace{-0.35cm}
		\end{subfigure}
		\begin{subfigure}{1\linewidth}
			\includegraphics[width=\textwidth]{"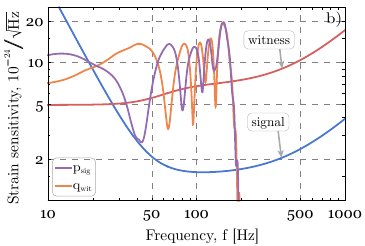"}
            \phantomsubcaption
            \label{fig:sc:PSD}
		\end{subfigure}
		\caption{\label{fig:scplot}Scattering signal from a fast scattering event. a) Normalized projections in the signal $\text{p}_{\text{sc}}$ and witness quadrature $\text{a}_{\text{sc}}$ of a scattering signal in time domain. b) Amplitude spectral density of the same scattering signal compared to the signal and witness quadrature sensitivities. Here, we can see the characteristic scatter shoulder produced by a fast scattering event.}
	\end{figure}

    The projections of a scattering signal in the signal quadrature $p_{\text{sig}}(t)$ and witness quadrature $q_{\text{wit}}(t)$ in time domain are given by~\cite{AstM2016,AstPhD2017}:
    \begin{equation}
		p_{\text{sig}}(t)=A\cos\bigl[\varphi_{0}+m\sin(2\pi f_{m}t+\phi_{m})\bigr],
		\label{eq:sc_model_phase}
	\end{equation}
     \begin{equation}
		q_{\text{wit}}(t)=A\sin\bigl[\varphi_{0}+m\sin(2\pi f_{m}t+\phi_{m})\bigr].
		\label{eq:sc_model_amp}
	\end{equation}
     $A$ is the amplitude of the scattering signal, $m$ is the modulation depth, $f_{m}$ is the modulation frequency, $\phi_{m}$ is the initial phase shift of the modulation and $\varphi_{0}$ is the initial phase shift of the projection in the two quadratures. This model describes fast scattering events, which produce so-called scatter shoulders. Frequency upconversion of a low frequency motion leads to a signal at higher frequencies. The maximum frequency component of a scatter shoulder scales linearly with the frequency and modulation depth of the scatter source and can be in the frequency range where detectors are most sensitive. Fig.~\ref{fig:sc:time} shows the normalized projections of a scattering signal in both quadratures in time domain for the model parameters: $A=2.13\times10^{-22}$, $f_{m}=5\,\text{Hz}$, $m=32.7$, $\phi_{m}=0$ and $\varphi_{0}=0$. To estimate the amplitude spectral density of these signal, we used Welch's method and applied a Savitzky-Golay filter to smoothen the resulting curves. The resulting spectral density of the scatter shoulder with a peak around $150\,\text{Hz}$ is shown in Fig.~\ref{fig:sc:PSD}. The plot also highlights that scattering signals can be so strong that we detect them in both quadratures.

    \subsection{Statistic}
    The SNRs of fast scattering events span a wide range. As a practical reference, we adopt the SNR distribution of all events (fast and slow) reported in~\cite{SoniReducingscatteredlight2021} and treat them as if they were all fast events. This choice is motivated by two considerations: (i) we lack a reliable model for slow scattering; (ii) using the full catalog significantly increases the sample size and yields more realistic SNR distributions to test our approach. Moreover, slow scattering events are typically louder and produce signals at frequencies lower than fast scattering events. Since the witness quadrature is most sensitive at low frequencies, slow events should be at least as detectable as the fast ones in our scheme. This makes our substitution a conservative: we do not make statements about the true composition of data statistics, but only demonstrate the potential of our approach to reduce the residual scattering.\\
    Our reference set, provided by Gravity Spy~\cite{BahaadiniGravitySpy2018}, contains all LIGO Livingston O3a events classified as slow or fast scattering with $10 \leq \mathrm{SNR} \leq 120$ and machine learning confidence $>0.95$. Our scattering waveform depends on multiple parameters whose marginal distributions are unknown. We therefore incorporate the reference SNR distribution only via the distribution of scattering amplitudes, keeping the other parameters fixed at: $f_{m}=5\,\text{Hz}$, $m=32.7$, $\phi_{m}=0$ and $\varphi_{0}=0$. We chose a relatively high modulation frequency $f_m$ to avoid a biasing the study towards the witness quadrature: increasing $f_m$ shifts more power to higher-frequency sidebands where the witness quadrature is less sensitive, resulting in a less favorable detectability (thus a harder test). In addition, we compute the spectral densities over multiple oscillations of the scattering signals, which averages out the impact of the initial phase and prevents accidental preference of one quadrature over another. 
    
\begin{figure}
 		\begin{subfigure}{1\linewidth}
			\includegraphics[width=\textwidth]{"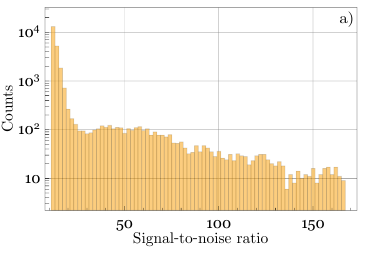"}
            \phantomsubcaption
            \vspace{-0.35cm}
            \label{fig:sc:hist:phase}
		\end{subfigure}
		\begin{subfigure}{1\linewidth}
			\includegraphics[width=\textwidth]{"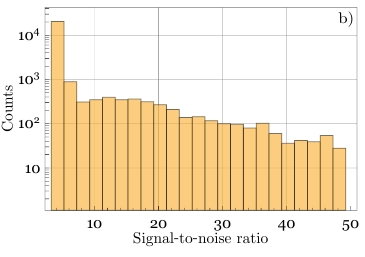"}
            \phantomsubcaption
            \label{fig:sc:hist:amp}
		\end{subfigure}
		\caption{\label{fig:sc:hist}Histograms of our set of of random scattering events. a) Histogram of signal-to-noise ratios for the signal quadrature. b) Histogram of signal-to-noise ratios for the witness quadrature.}
	\end{figure}

    Since our modeled sensitivity differed from the one used in a reference set for Advanced LIGO, we used the following protocol to ensure a realistic statistical distribution. From the reference data, we approximated the probability density function of the SNR distribution and used it to generate a set of random scattering amplitudes. This provided us with a realistic set of random scattering events, whose statistic would match the reference statistic if we had the sensitivity of Advanced LIGO for which the reference set was collected. \\
    Fig.~\ref{fig:sc:hist} shows the distribution of SNRs for our detector concept. We generated a set of 25000 random scattering events, each with at SNR of at least $8$ for the sensitivity of Advanced LIGO. Afterwards, we computed the SNRs in both quadratures for each event. In Fig.~\ref{fig:sc:hist:phase}, we present the distribution of SNRs for the signal quadrature. Since the sensitivity of the detector we consider here is higher than that of Advanced LIGO, corresponding SNRs are also higher compared to the reference. Fig.~\ref{fig:sc:hist:amp} shows the distribution of SNRs for the witness quadrature. As expected, they are smaller compared to those of the signal quadrature since the witness quadrature is less sensitive for most frequencies.

    \section{Results}
\begin{figure}
 		\begin{subfigure}{1\linewidth}
			\includegraphics[width=\textwidth]{"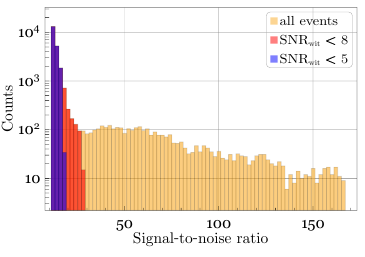"}
            \phantomsubcaption
                \vspace{-0.35cm}
		\end{subfigure}
		\begin{subfigure}{1\linewidth}
			\includegraphics[width=\textwidth]{"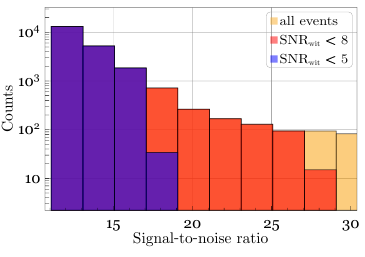"}
            \phantomsubcaption
		\end{subfigure}
		\caption{\label{fig:phasered}Histogram of signal-to-noise ratios for the signal quadrature. The histograms show the remaining scattering events after subtraction. The red events are the events that can not be subtracted if the witness quadrature threshold is set to $8$. Blue are the remaining events for which the signal-to-noise ratio in the witness quadrature is smaller than the threshold $5$. Top: Whole signal-to-noise ratio range. Bottom: Focus on the remaining scattering events.}
	\end{figure}

    In this section, we analyze the potential of our detector concept to filter scattering events out. The general idea is to detect the scattering signal in the witness quadrature. This information can then be used to reconstruct the scattering signal in the signal quadrature. Afterwards, the reconstructed scattering signal is subtracted from the signal quadrature data. This would increase the sensitivity to GW signals. On top of that, the witness quadrature channel makes it possible to veto scattering events. This method depends strongly on how well the scattering signal can be reconstructed. To take this problem into account, we considered two cases with different SNR thresholds: $8$ and $5$. We assumed that a scattering signal can be reconstructed well enough if the signal-to-noise ratio in the witness quadrature exceeds the threshold, although the exact reconstruction approach and its limitations are not considered here.\\
    In Fig.~\ref{fig:phasered} we show the statistic of our simulated scattering events before and after subtraction. In the case where we set the subtraction threshold to $8$, we can reduce the number of scattering events by $13.9\%$. In the other case, we can subtract $19.3\%$ of the scattering events. In both cases we are able to subtract all loud events, but the sensitivity is dominated by multiple quiet events, which we do not have sufficient sensitivity to.

\begin{figure}
		\includegraphics[width=3.375in]{"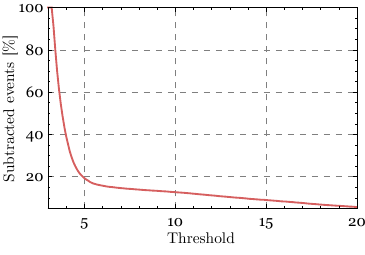"}
		\caption{\label{fig:phaseredper}Percentage of subtracted scattering events versus required signal-to-noise ratio in the witness quadrature. The plot highlights that the performance of our approach strongly dependence on the minimal signal-to-noise ration required for the reconstruction of the scattering signal. Especially below a threshold of $5$ we can see that the percentage of subtractable events increases drastically with decreasing threshold.}
	\end{figure}

    As mentioned previously, the potential to subtract scattering events depends on the choice of the signal-to-noise ratio threshold in the witness quadrature. Here, we considered rather high thresholds. As shown in Fig.~\ref{fig:phaseredper}, lowering the required signal-to-noise ratio in the witness quadrature could significantly improve the percentage of subtracted events. For a threshold of $4$, we could subtract $39.1\%$ of the events and even all events for a threshold of $3.2$.

\section{Summary and conclusion}
Previous research has shown that quantum-dense sensor technology is suitable for simultaneously reading out precise information in two non-commuting quadratures of the electromagnetic output field of an interferometer \cite{SteinlechnerS2013,AstM2016,Zander2021}. This quantum technology approach is highly relevant for GW detectors. While the GW signal appears exclusively in the signal quadrature of the output beam, non-stationary disturbances caused by backscattered laser light appear correlated in both outputs. The principle has been successfully demonstrated in tabletop experiments in the quantum measurement noise-dominated range. However, it is particularly relevant for GW detectors because the quantum technology approach can also be used at frequencies where quantum backaction noise rotates the optimal GW readout quadrature depending on the frequency. In this work, we propose for the first time an approach that is inspired by the quantum-dense metrology and allows to achieve frequency-dependent quantum enhancement simultaneous to reading out the orthogonal witness quadrature. While it does not take full advantage of quantum-dense metrology features, it has the advantage that it could be implemented in current GW detectors without increased quality requirements.\\
To evaluate the potential of our approach in the backaction regime, we generated a set of random scattering events and analyzed how many of these events can be subtracted. It showed that we can filter the very loud scattering events out. The actual performance strongly depends on the signal-to-noise ratio needed in the witness quadrature to reconstruct the signal. Using tools such as machine learning could lower the requirement and significantly improve the performance of our approach.\\
    The additional quadrature-witness readout comes at a price of a reduced signal quadrature sensitivity at high frequencies. In the case of equally strong quadrature detection, this corresponds to a loss in GW sensitivity of approximately $3\,\text{dB}$. This loss can be minimized by optimizing the properties of the readout beamsplitter and the second squeezer. A more detailed optimization requires further insight into the reconstruction of the scattering signals. Alternatively, a filter cavity could be used to separate the signals so that low-frequency components are measured by dual readout, while high-frequencies are measured by a traditional single readout~\cite{MeindersScattering2015}.\\

    Our concept allows to mitigate loud back-scattering events without modification of the core layout of the detector. Crucially, this approach is not limited to back-scatter noise: other noise that produces a signal in the orthogonal quadrature, e.g.~technical noise, could be reduced in the same way. 

	\begin{acknowledgments}
    This work was supported by the Deutsche Forschungsgemeinschaft (DFG) under Germany’s Excellence Strategy EXC 2121 “Quantum Universe”-390833306.
	\end{acknowledgments}
    
    \bibliography{Niels_Bib}
    
\end{document}